# Lithography-Free Fabrication of Graphene Devices

N. Staley[1], H. Wang[1], C. Puls[1], J. Forster[1], T.N. Jackson[2], K. McCarthy[1], B. Clouser[1], and Y. Liu[1,†]
[1]Department of Physics, Pennsylvania State University, University Park, PA 16802
[2]Department of Electrical Engineering, Pennsylvania State University, University Park, PA 16802
[†]e-mail address: liu@phys.psu.edu

## Abstract

We have developed a lithography-free, all-dry process for fabricating graphene devices using an ultrathin quartz filament as a shadow mask to avoid possible contamination of graphene during lithographic process. This technique was used to prepare devices for electrical transport as well as planar tunnel junction studies of *n*-layer graphene (nLG), with $n = 1, 2, 3$ and higher. We observed localization behavior and an apparent reduction of density of states (DOS) near the Fermi energy in nLG.

There has been a flurry of recent work[1,2] on films of 1-layer graphene (1LG), motivated by the pioneering work of Geim and coworkers[3] and Kim and coworkers[4]. Surprisingly, 1LG was found to host a two-dimensional electron gas with a band structure featuring zero effective mass[5]. Two types of unconventional integer quantum Hall effects (IQHE) were observed in 1LG[1,2] and in 2-layer graphene (2LG)[6] devices, respectively. Theoretical calculation indicates that *n*-layer graphene (nLG) with $n > 2$ are also interesting[5].

The highest mobility reported for 1LG devices is around 10,000 $cm^2/Vs$ at high gate voltages[2], which is remarkable. However, it may not be sufficiently high to allow the observation of certain physical phenomena, such as fractional quantum Hall effect (FQHE). So far, all graphene devices reported in the literature were prepared by e-beam lithography. Multiple steps are required to pattern a device, including coating with organic materials, which may subject the graphene to possible contamination and add unwanted disorder to the device. It is therefore desirable to pursue alternative graphene device fabrication. Using ultrathin quartz filaments as shadow masks, we have developed a method to fabricate graphene devices, aiming at raising the mobility of the devices. Our method is lithography-free, all dry, and simple to implement. Devices fabricated were measured using a DC technique with a typical excitation current of 1 μA in a dip probe in which the sample was cooled by direct contact with $^4$He liquid or gas.

Two methods have been used to create graphene samples - exfoliation either mechanically in air[7] or chemically in solutions[8], and thermal decomposition of SiC[9]. Our nLG flakes were created by mechanical exfoliation in air from freshly cleaved highly oriented pyrolytic graphite (HOPG)[10]. Heavily *N*-doped silicon with a 300-nm-thick thermally grown $SiO_2$ top layer was used as substrates. Thin graphene flakes were identified under an optical microscope with 500X magnification, as reported previously[1]. After inspecting a large number of exfoliated graphene flakes we concluded that the color and the faintness of the optical images of exfoliated graphene flakes fall into distinguishable patterns that can be organized into a "color code" scheme for nLG. This color-code scheme was correlated with AFM measurements in which the minimal height at the edge of the graphene flake was taken as the flake thickness. We also attempted to correlate our color-code scheme with the Raman spectroscopy measurements that provide information on the thickness of graphene flakes[11].

An ultrathin quartz filament as thin as 200 nm in diameter, pulled from pure quartz melt as described previously[12], was placed onto the graphene flake of interest and used as a shadow mask. It was found that filaments around 1 μm in diameter could be manipulated most easily on a $SiO_2$ surface. A film of Au was then evaporated, creating two electrodes serving as the source and drain (Figs. 1a and b). The heavily doped Si substrate was used as a back gate. We also fabricated planar tunnel junctions by evaporating a 2-nm-thick Al film at an angle ($45^o$) that was allowed to oxidize in air to form an $Al_2O_3$ tunnel barrier (Fig. 1c). Two electrodes of Pb or Au were then deposited at a different angle, either vertical or $135^o$, to form a tunnel junction (Fig. 1d). In both cases two fine Au wires were attached to each electrode using Ag epoxy to allow linear conductivity or tunneling measurements. Simultaneous measurements of both linear and Hall conductivities can be achieved by using a second filament to form a cross mask, which results in a van der Pauw probe[13].

In Fig. 2a, we show the conductance ($\sigma$) verses temperature ($T$) for a 1LG device at three gate voltages ($V_g$). Here $\sigma$ is calculated from sample resistance per square, $R_\square$, with the contact resistance ($R_c$) included. It was reported previously[4] that the contact resistance ($R_c$) between Au and graphene is small. The value of $R_c$, which could be inferred from examining the sample resistance at high $V_g$ (providing an upper bound), or the linearity in the $\sigma$ *vs.* $V_g$ plot, was found to be much less than $R_\square$ for our nLG devices for small $n$. The hysteresis in $\sigma$ *vs.* $V_g$ is seen to be small (Fig. 2b), suggesting that few interface charge traps were present at or near the graphene-$SiO_2$ interface in this device. It is seen that this 1LG device follows a $\sigma \propto \ln(T)$ behavior for 8 K < $T$ < 50 K, consistent with weak localization in two dimensions (2D)[14]. It was reported previously that quantum interference responsible for weak localization behavior could be suppressed in 1LG because of the local deformation of 1LG that results in random effective magnetic fields[15]. However, weak localization behavior was found at least in one sample[15]. It was emphasized that for 1LG, a minimum conductivity of $4e^2/h$ was found at the Dirac point[1]. In other studies, however, conductivities very different from $4e^2/h$ were also observed[2,4]. Interestingly, for this particular 1LG sample we measured, $\sigma$ appears to level off below 8 K at values close to $4e^2/h$ (= 1.55 mS). The implication of this observation is to be understood.

In Fig. 3a, $\sigma$ *vs.* $V_g$ of a gated 5LG device is shown. The asymmetric response suggests that the conduction and valence bands in a 5LG system have significant overlap, which leads to a density of states (DOS) that is asymmetric with respect to the Fermi energy, $E_F$, consistent with theoretical expectations[5]. For a device with single type of carrier (electrons or holes), $\mu = \sigma/n_c e$, where $\mu$ is the mobility, $\sigma$ the conductivity, $n_c$ the carrier density, and e the elemental charge. Because of the lack of Hall measurements, only the change of the carrier density by ramping $V_g$ can be calculated from the capacitance. The mobility due to field effect is $\mu_{FE} = (d/\varepsilon_0\varepsilon)\ (\partial\sigma/\partial V_g)$, where $d$ is the thickness of the insulating barrier, $\varepsilon_0$ the permittivity of free space, and $\varepsilon$ the dielectric constant. For this 5LG device, $\mu_{FE}$ = 6,700 $cm^2$/Vs for the holes at high (negative) gate voltages, which is not as high as that previously reported for 1LG devices. Refinement of our lithography-free technique will be needed to achieve higher mobility.

In this 2D 5LG device (1.8 nm thick), a temperature range in which $\sigma$ shows *ln(T)* behavior can be easily identified at $V_g$ = 0 and 50 V (Figs. 3b and c). This *ln(T)* behavior may be associated with either quantum interference in the coherent back scattering or interaction effects for diffusive 2D carriers[14]. The observation of the *ln(T)* behavior up to 50 K was surprising since at such high temperatures, the dephasing length should be very small. Interestingly, $V_g$ = 10 V, close to the minimum in $\sigma(V_g)$, $\sigma$ is seen to follow the *ln(T)* behavior less precisely (Fig. 3d). This is reasonable - As the carrier density is lowered, the system becomes effectively more strongly disordered, leading eventually to deviation from the weak localization behavior.

In Fig. 4, we show results on a planar Pb-$Al_2O_3$-6LG tunnel junction. Pb was chosen to be the top electrode primarily to test the fabrication process. As shown in Fig. 4a, the junction resistance, $R_J(T)$, shows insulating behavior, suggesting that we do have a good tunnel barrier. A resistance drop was found at 6.8 K (Fig. 4a), the $T_c$ of the Pb electrode, which is slightly lower than that of bulk Pb (7.2 K). At 4.2 K, below the $T_c$ of the Pb electrode, the tunnel conductance, proportional to DOS, showed a drop around 2 meV, suggesting the opening of an energy gap (Fig. 4b). The gap value was roughly twice of what is expected for Pb, most likely because two tunnel junctions were formed in series, one with the designed $Al_2O_3$ barrier and the other formed unintentionally because of the oxidation of Pb. There appears to be a suppression of DOS even at 10 K, above the $T_c$ of Pb. Preliminary results obtained on tunnel junctions of Au-$Al_2O_3$-nLG suggest the same suppression.

We would like to acknowledge the useful discussions with Profs. Jun Zhu, Jainendra Jain, Jorge Sofo, and Peter Eklund, and Drs. Paul Lammert and Paul Campbell, and technical assistance from Dalong Zhao and Awnish Gupta. We would also like to thank Profs. Theresa Meyer and Tom Mallouk for the use of their optical microscopes. This work is support in part by DOE under grant DE-FG02-04ER46159.

## Figures and captions

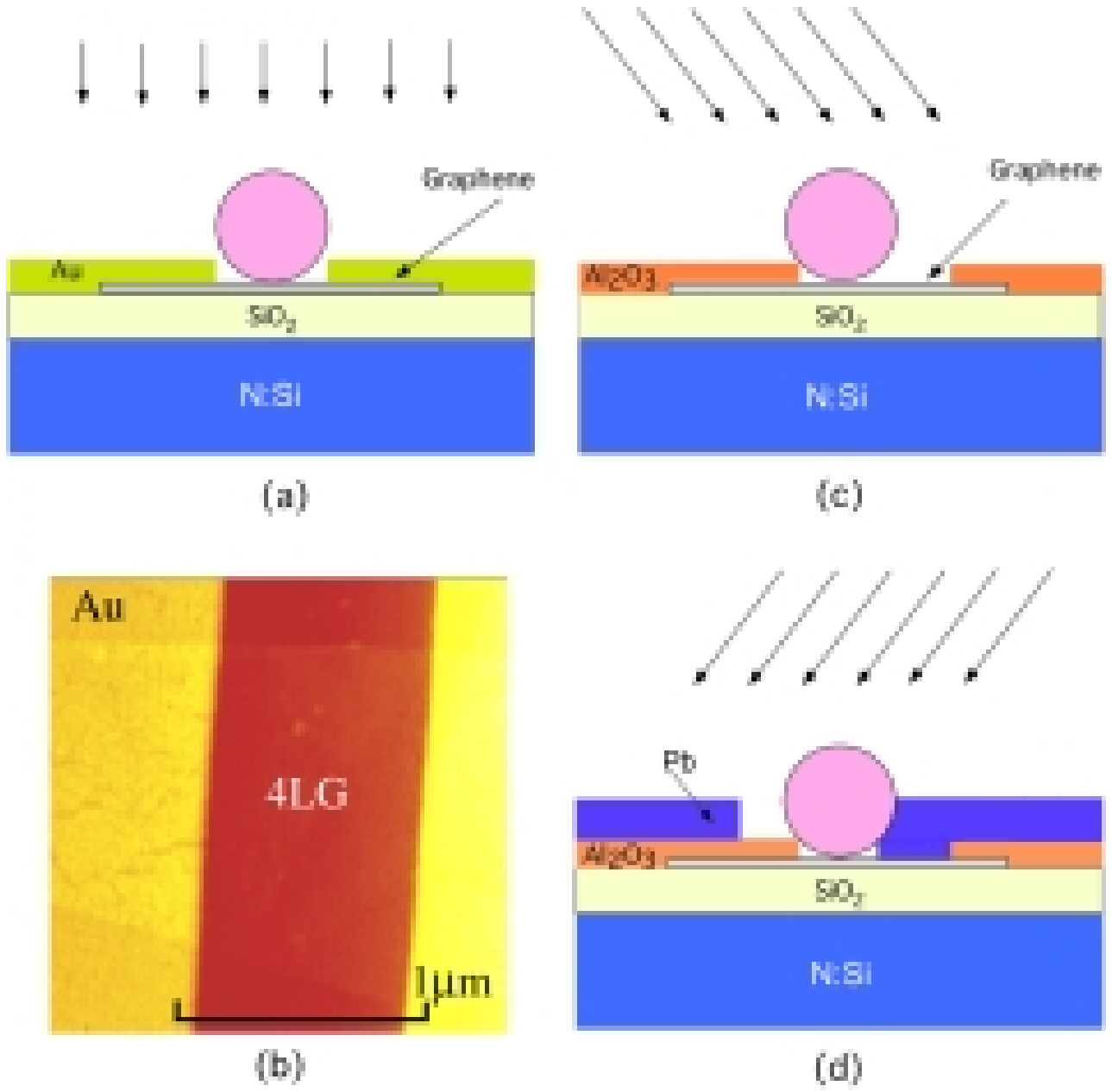

FIG. 1 (Color online): a) Schematic for nLG device fabrication for transport measurement. An ultrathin quartz filament is used as a shadow mask; b) An AFM image of an actual 4LG device. The graphene flake is visible; c-d) Schematics for planar tunnel junction fabrication.

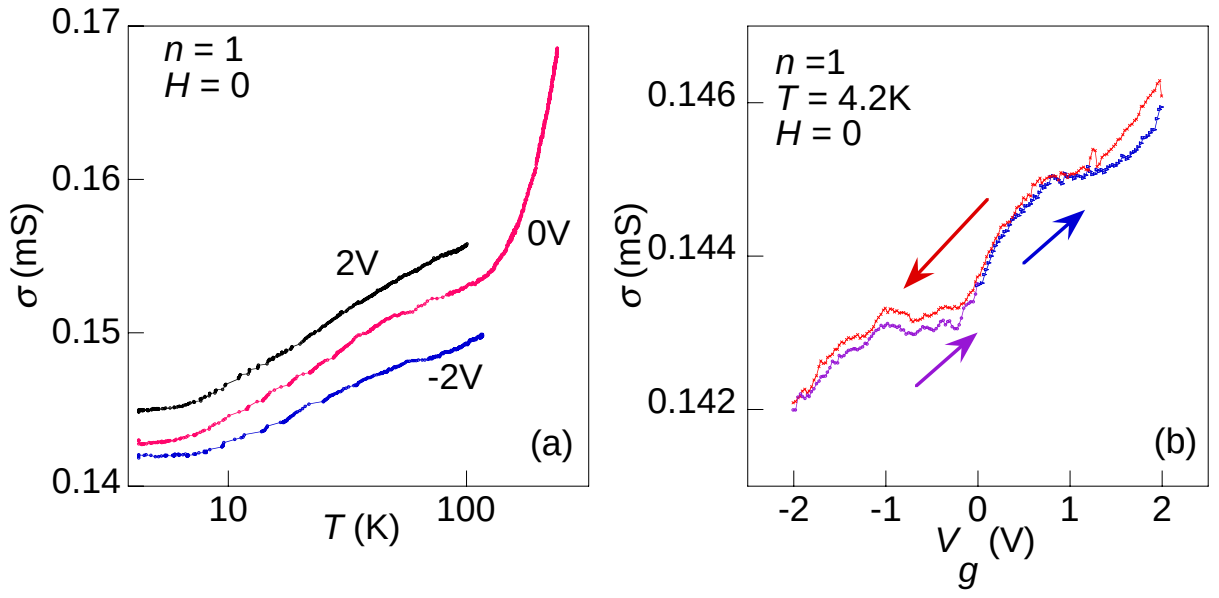

FIG. 2: a) $\sigma$ *vs.* $T$ for a 1LG device at three $V_{\rm g}$ values. b) $\sigma$ *vs.* $V_{\rm g}$ loop. The arrows indicate how $V_{\rm g}$ was swept. Small hysteresis is seen. The range of $V_{\rm g}$ was limited in this device because of a leakage created by Ag epoxy spilled over to the edge of the doped Si substrate used as the back gate.

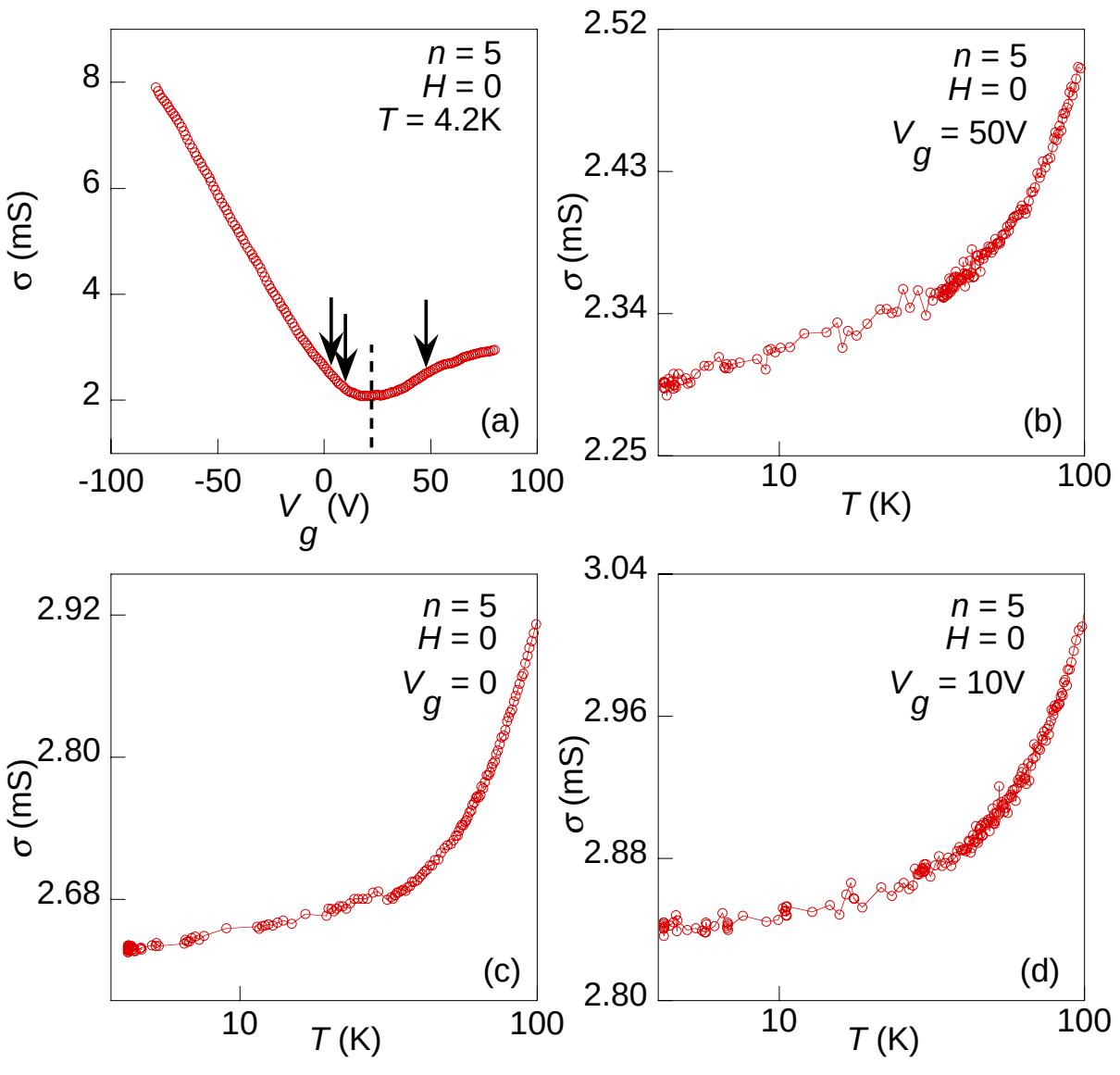

FIG. 3: a) $\sigma$ *vs.* $V_g$ for a 5LG device. The arrows indicate the $V_g$'s at which $\sigma$ *vs.* $T$ measurements were taken and the dash line shows the $V_g$ at which $\sigma$ reaches a minimum (25 V); b-d) $\sigma$ *vs.* $T$ in semilog plot at various $V_g$ values as indicated.

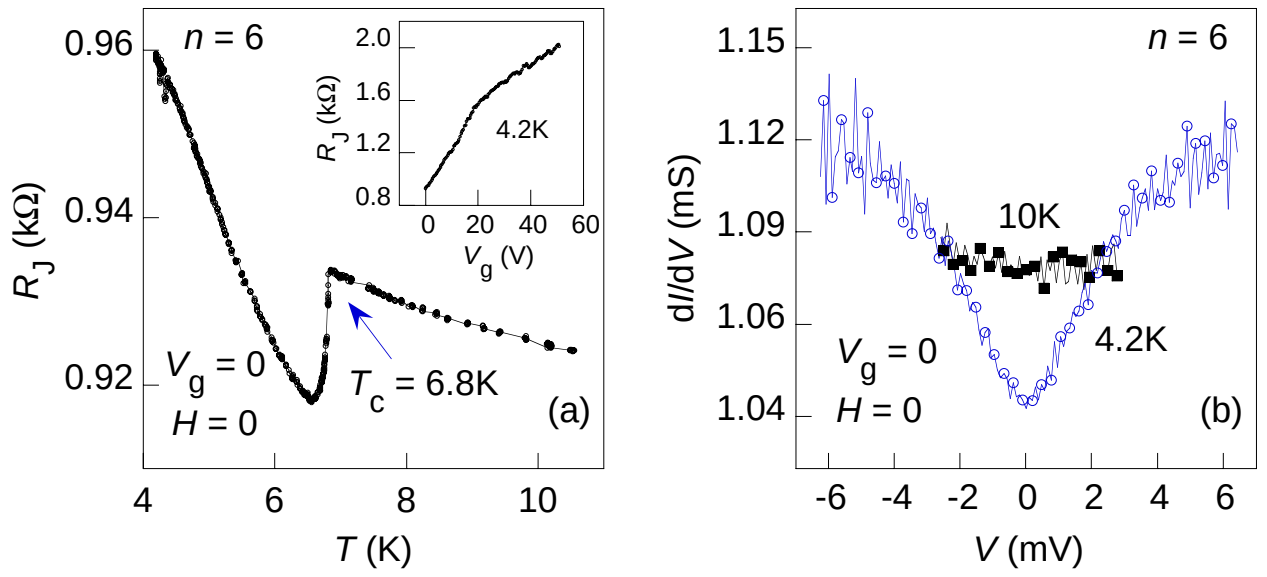

FIG. 4: a) $R_J$ *vs.* $T$ for a Pb-$Al_2O_3$-6LG planar tunnel junction. Insert: $R_J$ *vs.* $V_g$ at 4.2 K; b) d*I*/d*V* *vs.* *V* at two *T*s as indicated. The drop in d*I*/d*V* near 2 meV at 4.2 K indicates the opening of an energy gap. Data on d*I*/d*V* at voltages higher than 2 meV at 10 K is not available.

---